\documentclass[journal]{IEEEtran}

\usepackage{cite}
\usepackage{amsmath,amssymb,amsfonts,amsthm}
\usepackage{algorithmic}
\usepackage{graphicx}
\usepackage{textcomp}
\usepackage{xcolor}
\usepackage{hyperref}
\usepackage[ruled]{algorithm2e}
\usepackage{autobreak}
\usepackage{multirow}
\hypersetup{hidelinks}
\newtheorem{remark}{Remark}

\def\Pb{\mathbf{P}}
\def\pb{\mathbf{p}}

\def\qb{\mathbf{q}}

\def\Xb{\mathbf{X}}
\def\xb{\mathbf{x}}

\def\Bm{\boldsymbol{\mathcal{B}}}

\def\Em{\boldsymbol{\mathcal{E}}}
\def\Fm{\boldsymbol{\mathcal{F}}}

\def\Pm{\boldsymbol{\mathcal{P}}}
\def\Qm{\boldsymbol{\mathcal{Q}}}
\def\Mm{\boldsymbol{\mathcal{M}}}
\def\Sm{\boldsymbol{\mathcal{S}}}

\def\Xm{\boldsymbol{\mathcal{X}}}

\def\Pcal{\mathcal{P}}
\def\Rcal{\mathcal{R}}

\begin{document}

\title{DF-3DRME: A Data-Friendly Learning Framework for 3D Radio Map Estimation based on Super-Resolution Technique}
\author{Lin~Zhu,
        Weifeng~Zhu,
        Shuowen~Zhang,
        Giuseppe~Caire,
        and~Liang~Liu% <-this % stops a space
\thanks{L. Zhu, W. Zhu, S. Zhang, and L. Liu are with the Department of Electrical and Electronic Engineering, The Hong Kong Polytechnic University, Hong Kong SAR (email: \href{lin-eee.zhu@connect.polyu.hk}{lin-eee.zhu@connect.polyu.hk}, \{\href{eee-wf.zhu@polyu.edu.hk}{eee-wf.zhu}, \href{shuowen.zhang@polyu.edu.hk}{shuowen.zhang}, \href{liang-eie.liu@polyu.edu.hk}{liang-eie.liu}\}@polyu.edu.hk).}
\thanks{G. Caire is with the Communications and Information Theory Group, Technische Universit{\"a}t Berlin, Berlin, Germany (e-mail: \href{caire@tuberlin.de}{caire@tuberlin.de}).}
\thanks{An earlier version of this paper was presented in part at the 2026 IEEE International Conference on Communications (ICC) Workshops~\cite{DF-3DRME_Conf}.}
}% <-this % stops a space}
\maketitle

\begin{abstract}
High-Resolution three-dimensional (3D) radio maps (RMs) provide rich information about the radio landscape that is essential to a myriad of wireless applications in the future wireless networks. Although deep learning (DL) methods have shown their effectiveness in RM construction, existing approaches require massive high-resolution 3D RM samples in the training dataset, the acquisition of which is labor-intensive and time-consuming in practice. In this paper, our goal is to devise a \emph{data-friendly} high-resolution 3D RM construction solution via training over a \emph{hybrid} dataset, wherein the RMs associated with a small fraction of environment maps (EMs) are of high-resolution, while those corresponding to the majority of EMs are of low-resolution. To this end, we propose a \underline{D}ata-\underline{F}riendly \underline{3D} \underline{R}adio \underline{M}ap \underline{E}stimator (DF-3DRME), which comprises two processing stages. Specifically, in the first stage, we leverage the abundant low-resolution 3D RM samples to train a neural network, termed the LR-Net, for predicting the low-resolution 3D RM from the input EM, which provides a coarse characterization of the spatial radio propagation. In the second stage, we employ an advanced \emph{super-resolution} network, termed the SR-Net, to upscale the predicted low-resolution 3D RM to its high-resolution counterpart. Unlike the LR-Net, the SR-Net can be effectively trained with only the limited high-resolution 3D RM samples available in the hybrid dataset. Experimental results demonstrate that the proposed framework achieves compelling reconstruction performance with only 4\% of the EMs in the dataset having high-resolution 3D RM labels, which significantly reduces data acquisition overhead and facilitates practical deployment.
\end{abstract}


\IEEEpeerreviewmaketitle

\vspace{-0.2cm}
\section{Introduction}
%\IEEEPARstart{F}{or} 
Accurate characterization of radio propagation in wireless environment is fundamental to the design and optimization of cellular systems. Conventional approaches to radio propagation modeling predominantly rely on statistical models, which parameterize the channel as a stochastic function of variables such as the transmitter-receiver distance~\cite{Molisch}. For example, the distance-dependent path loss model with log-normal shadowing is widely adopted in traditional cellular systems~\cite{Rappaport, Goldsmith}. However, these statistical models offer only coarse descriptions of the propagation channel and cannot capture the fine-grained channel variations induced by site-specific structures for wireless networks. Recently, radio maps (RMs) have emerged as a principled framework to overcome the limitation~\cite{SongyangZhangMagzine, SuzhiBiTWC, DanielSPM, ZengYongTutor}, which can accurately capture the site-specific radio propagation landscape of the environment. By providing position- and environment-specific channel information, RMs enable a wide range of wireless applications, including spectrum reuse in cognitive networks~\cite{SuzhiBiTWC, DanielSPM}, localization~\cite{LocUNet}, and unmanned aerial vehicle (UAV) path planning~\cite{ShuowenZhangUAVTWC, UAVPathPlan}, and are therefore regarded as an important enabler of the environment-aware communication paradigm envisioned for sixth-generation (6G) systems and beyond.

Traditional cellular systems are mainly responsible to serve the terrestrial users, thus merely necessitating the two-dimensional (2D) RM at the ground level. However, the emergence of the low-altitude economy~\cite{LAE} and space-air-ground integrated networks~\cite{SAGIN} demands three-dimensional (3D) RMs capable of characterizing channel conditions throughout an expansive 3D space. Furthermore, the spatial resolution of RMs is also of paramount importance, because a high-resolution RM can provide fine-grained channel information that significantly improves the performance of communication and sensing tasks in wireless networks. For example, the resolution of the RM determines the localization accuracy of the scheme proposed in~\cite{LocUNet}. Therefore, it is crucial to develop efficient methods for constructing high-resolution 3D RMs.

However, in cellular networks, obtaining these maps solely through real-world measurements is labor-intensive and time-consuming, as it requires extensive site surveys that densely sample the entire 3D space. To tackle this data collection issue, data-driven interpolation methods and model-based data fitting methods have been proposed~\cite{IDW, Kriging, Kernel, FuXiaoTSP, SunHaoTSP, SLF}. These approaches only need to obtain channel measurements at sparsely distributed locations and adopt a series of parametric models to infer the channel conditions at unmeasured positions. However, such methods are predominantly applied to open environments with simple propagation model, and exhibit poor performance in complex and irregular setups such as urban areas and indoor scenarios.

A fundamentally different strategy is to leverage the environment map (EM), i.e., the geometric layout of buildings and terrain features, to predict the RM. This approach is grounded in the physical principle that, given the EM and the transmitter configuration, the RM is uniquely determined by Maxwell's equations~\cite{Molisch, RTcost}. Ray-tracing simulators can approximate this physical process and produce accurate RMs, but their computational cost becomes prohibitive at high spatial resolutions over large 3D domains~\cite{Lecci2021, RTcost}. More recently, deep learning (DL) techniques have been shown to learn the complex mapping from EMs to RMs, which effectively approximates ray-tracing with significantly reduced cost~\cite{RadioUNet, RMEGAN, RadioGAT, RadioDiff, RadioDiff3DSlice, SpectrumNet, SliceGNN, TianyuTVT, ZhiboLett, RadioDiff3D}. In this paper, we address the problem of constructing high-resolution 3D RMs by leveraging the powerful representational capabilities of DL.

\vspace{-0.3cm}
\subsection{Prior Works}
Recently, DL-based RM construction has emerged as a promising paradigm for environment-aware communications. In particular, \cite{RadioUNet} formulates the mapping from an EM to its corresponding RM as an image-to-image translation problem, demonstrating that a U-Net architecture~\cite{UNet} can yield accurate predictions when sufficient labeled data are available. Building upon this methodology, subsequent works have adopted generative adversarial network (GAN)~\cite{RMEGAN}, graph neural network (GNN)~\cite{RadioGAT}, and diffusion model~\cite{RadioDiff} to further enhance the reconstruction accuracy. Compared with data-driven interpolation and model-based data fitting methods, DL-based approaches can capture highly nonlinear relationships between environment representations and RMs, and thus provide more accurate RM estimates in practical environments.

Although the aforementioned DL-based methods have demonstrated remarkable effectiveness, they are primarily designed for 2D RM prediction. To extend the DL paradigm to 3D scenarios, several studies have proposed constructing 3D RMs by predicting a stack of 2D RMs sliced at discrete altitudes~\cite{RadioDiff3DSlice, SpectrumNet, SliceGNN}. For instance, diffusion models~\cite{RadioDiff3DSlice}, U-Net-based predictors~\cite{SpectrumNet}, and GNN-based architectures~\cite{SliceGNN} have been employed to first generate a 2D RM at each altitude and then assemble these 2D slices into a 3D RM.  However, such slice-based methods exhibit a fundamental limitation of neglecting the correlation along the vertical dimension for the 3D RM due to the 3D wave propagation. As a result, these methods usually suffer from inconsistent shadow boundaries across adjacent altitudes in the reconstructed 3D RM, leading to degraded reconstruction quality.

To better capture the 3D spatial dependence, recent works have begun exploring the construction of \emph{generalized} 3D RM, where the channel information at all grids of the 3D space are directly generated by a 3D neural network~\cite{TianyuTVT, ZhiboLett, RadioDiff3D}. Such architectures can inherently capture spatial correlations across all three dimensions, yielding more physically consistent reconstructions. For instance, \cite{TianyuTVT} integrates GAN-based learning to reconstruct 3D RMs from measurements collected along a UAV trajectory, while \cite{ZhiboLett} further optimizes the UAV trajectory and employs a 3D convolutional autoencoder to enhance reconstruction performance. However, these works consider RM construction with low spatial resolutions, which lacks the fine-grained propagation details in dense urban environments. More recently, \cite{RadioDiff3D} proposes a 3D diffusion model that takes both EMs and measurements as inputs to generate the complete 3D RMs. Although this approach realizes generalized 3D RM construction with relatively high spatial resolution, it only considers the 3D RM with an altitude span of 4 meters, which is substantially lower than the typical operating altitudes of UAVs and other aircraft. More critically, the 3D models proposed in~\cite{TianyuTVT, ZhiboLett, RadioDiff3D} share two common limitations that hinder the extension to generate high-resolution 3D RMs with expanded altitude coverage. First, they rely on auxiliary measurement inputs as a prerequisite for RM generation. This requirement becomes particularly burdensome for high-resolution 3D RMs, where even a modest sampling ratio can lead to an intractable number of measurement points in practice (e.g., the sampling ratios adopted in~\cite{RadioDiff3D}). Second, and more fundamentally, training these models demand a large-scale dataset comprising massive pairs of 3D EMs and the corresponding high-resolution 3D RMs, the acquisition of which incurs prohibitive computational and labor costs. 

\vspace{-0.4cm}
\subsection{Main Contributions}
In this paper, we investigate the high-resolution 3D RM reconstruction problem under a \emph{data-friendly} design philosophy. For the training phase, given the practical constraints associated with acquiring high-resolution 3D RM labels, we adopt a \emph{hybrid} dataset in which all the 3D EMs in the dataset possess low-resolution 3D RM labels, while only a small fraction of 3D EMs have the corresponding high-resolution 3D RM labels. For the inference phase, our framework takes only the 3D EM and the transmitter location as inputs to guarantee measurement-free inference. To effectively realize these objectives, we develop a \underline{D}ata-\underline{F}riendly \underline{3D} \underline{R}adio \underline{M}ap \underline{E}stimator (DF-3DRME), together with a multi-phase training strategy. The main contributions of this work are summarized as follows:
\begin{itemize}
  \item \textbf{Two-stage DL architecture for high-resolution 3D RM construction:} The proposed DF-3DRME is designed as a two-stage architecture to fully exploit the hybrid dataset while eliminating the reliance on auxiliary measurements. In the first stage, a compact neural network termed \textit{LR-Net} is proposed to directly predict the low-resolution 3D RM from the 3D EM and the transmitter location. In the second stage, a super-resolution network termed \textit{SR-Net} is employed to upscale the predicted low-resolution 3D RM to its high-resolution counterpart. Such a two-stage framework can overcome the difficulties in collecting a large number of high-resolution 3D RM samples during training, and enable measurement-free inference, making high-resolution 3D RM construction available for a wide range of environments.

  \item \textbf{Perceptual loss enhancement with multi-phase training strategy:} To enhance reconstruction quality, we design a composite loss function incorporating perceptual regularization, which effectively preserves structural features in the reconstructed high-resolution RMs. Furthermore, to fully exploit the hybrid dataset, we propose a multi-phase training strategy that progressively optimizes the two-stage network. Specifically, the LR-Net and SR-Net are first pretrained individually using the low-resolution and high-resolution 3D RM samples, respectively, followed by a fine-tuning phase that bridges the two stages to guarantee the end-to-end compatibility. This design can boost the representational capabilities of these two sub-networks while stabilizing the overall training process.

  \item \textbf{Hybrid dataset construction and comprehensive performance evaluation:} We construct a hybrid dataset for 3D RM construction and its extended researches, which is generated using realistic 3D EMs by ray-tracing (RT) simulations.\footnote{https://github.com/LINZHU-PolyU/Hybrid-3D-RM-Dataset} Based on this dataset, we conduct extensive experiments to validate the effectiveness of the proposed approach. The experimental results demonstrate that the proposed DF-3DRME significantly outperforms baseline approaches in terms of both numerical metrics and visual quality. More importantly, the results reveal that the proposed framework achieves compelling reconstruction performance when only 4\% of the EMs in the dataset possess high-resolution 3D RM labels, validating its data-friendly nature in practical deployment. 
\end{itemize}

\vspace{-0.3cm}
\subsection{Organization}
The remainder of this paper is organized as follows. Section~\ref{Sec: Def. of RM} presents the system model and the definition of 3D RM. Section~\ref{Sec: Hybrid Dataset} describes the structure of the considered hybrid dataset. Section~\ref{Sec: Two-stage} details the proposed two-stage model architecture together with the training methodology. Section~\ref{Sec: Experiments} presents the experimental results and analysis. Finally, Section~\ref{Sec: Conclusion} concludes this paper.

\textit{Notation}: Throughout this paper, $x$, $\xb$, $\Xb$, and $\boldsymbol{\mathcal{X}}$ denote a scalar, a vector, a matrix, and a tensor, respectively. The notation $\Xm(i, j, k)$ represents the $(i, j, k)$-th entry of a third-order tensor $\Xm$. The symbol $\| \cdot \|_1$ and $\| \cdot \|_F$ denote the $\ell_1$ norm and Frobenius norm for a matrix or a tensor, respectively. Finally, $[I] = \{1, 2, \cdots, I\}$ for any positive integer $I$.

\section{System Model And Definition of 3D RM} \label{Sec: Def. of RM}
\begin{figure*}[t]
    \centerline{\includegraphics[width=0.8\textwidth]{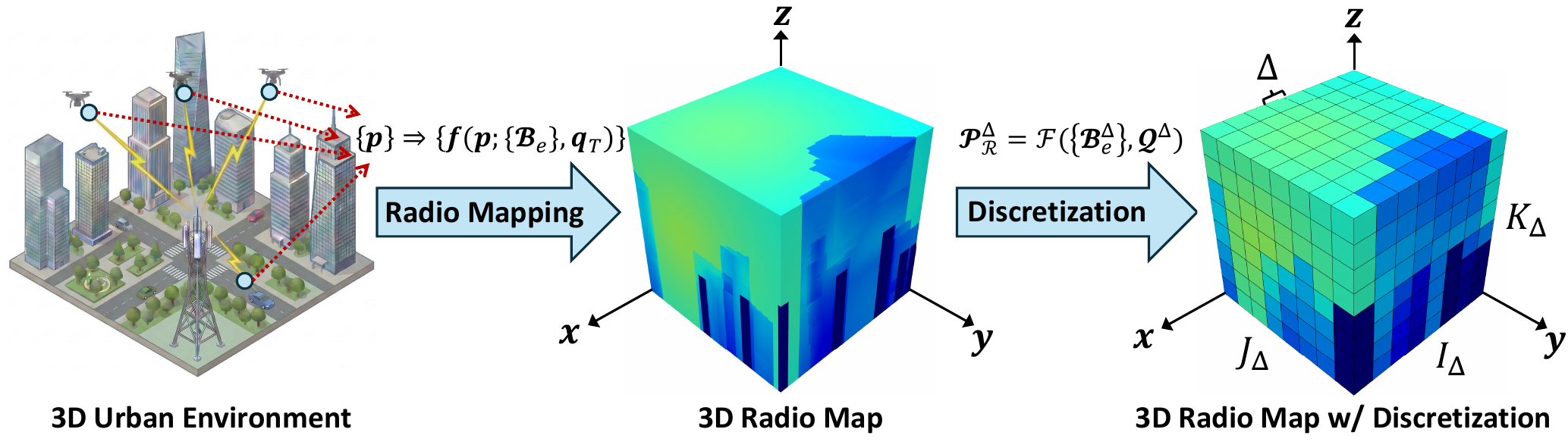}}
    \vspace{-0.3cm}
    \caption{Illustration of a 3D urban environment and the corresponding 3D RM discretized at resolution $\Delta$.}
    \label{Fig: 3D Environment}
    \vspace{-0.5cm}
\end{figure*}
\subsection{Path Loss Model and Propagation Environment} \label{Subsec: Channel Model}
Consider a 3D region of interest $\mathcal{R} \subset \mathbb{R}^3$. Let $\qb_T \in \mathcal{R}$ denote the location of a single-antenna transmitter that radiates a signal of power $P_t$. For a single-antenna receiver at location $\pb \in \Rcal$, let $P_r(\pb)$ denote the received signal power averaged over small-scale fading and shadowing. The \emph{path loss} at $\pb$ is then defined as the ratio of the transmit power to the average received power~\cite{Goldsmith}:
\begin{align} \label{Eq: PL}
    P_L(\pb) = \dfrac{P_t}{P_r(\pb)}.
\end{align}

Characterizing path loss has been a long-standing objective in wireless communications. Classical analytical models, such as the free-space path loss model, express $P_L(\pb)$ as a closed-form function of the distance between $\qb_T$ and $\pb$, neglecting the underlying propagation environment. Empirical models further incorporate statistical corrections derived from field measurements to account for real-world propagation effects. Representative examples include the 3GPP and WINNER propagation models~\cite{Goldsmith}, which express the path loss in dB scale as
\begin{align} \label{Eq: Empirical Path Loss}
    P_L(\pb)~{\rm dBm} = L_{f_c} + 10\gamma\log_{10}(d_{\pb}) + \kappa,
\end{align}
where $L_{f_c}$ is a constant capturing the frequency-dependent loss, $d_{\pb}$ is the distance between $\qb_T$ and $\pb$, $\gamma$ is the path loss exponent, and $\kappa$ is an environment-specific correction term. However, since $\gamma$ and $\kappa$ are typically calibrated via measurement campaigns based on coarse environmental classifications (e.g., urban, suburban, or rural), their applicability to heterogeneous deployments is limited \cite{RTcost}. More fundamentally, these models remain range-based. They assign identical path loss values to equidistant receivers regardless of their local surroundings, which is inconsistent with real-world observations.

To obtain a more accurate path loss characterization, we adopt an environment-aware and location-specific formulation. For a fixed transmit power, the average received power $P_r(\pb)$ is in principle governed by the propagation environment and the transmitter location $\qb_T$~\cite{Goldsmith}. Let $\mathcal{O} \subset \mathcal{R}$ denote the subregion occupied by physical obstacles (e.g., buildings, terrain, and foliage). We define the \emph{environment descriptor}, $\mathcal{B}$, as a set of spatially annotated points:
\begin{align} \label{Eq: Define Bcal}
\mathcal{B} = \left\{ (\pb,\, c_{\pb}) \mid \pb \in \Rcal \right\},
\end{align}
where $c_{\pb} \in \{0, 1\}$ is a binary indicator such that $c_{\pb} = 1$ if $\pb \in \mathcal{O}$, and $c_{\pb} = 0$ otherwise. Given this characterization, the path loss in~\eqref{Eq: PL} can thus be generalized as a deterministic function of $\mathcal{B}$ and $\qb_T$:
\begin{align} \label{Eq: PL function}
    P_L(\pb) = f\left(\pb;\, \mathcal{B}, \, \qb_T\right),
\end{align}
where $f(\cdot)$ denotes the propagation function that maps the receiver location, the environment descriptor, and the transmitter location to the corresponding path loss value. In practice, however, the true mapping $f(\cdot)$ is highly intricate due to complex propagation phenomena such as diffraction, scattering, and multiple reflections. This motivates a data-driven approach in which $f(\cdot)$ is approximated by a deep neural network, as detailed in Section~\ref{Sec: Two-stage}.

\subsection{Definition of 3D RM} \label{Subsec: 3D Radio Map}
Building on the path loss formulation in~\eqref{Eq: PL function}, we formally define the 3D RM, which refers to the spatial distribution of the path loss over the 3D space $\Rcal$. However, it is infeasible to store the path loss at each position in the environment due to finite storage\cite{UAVPathPlan, RadioUNet}. To enable efficient representation, we discretize $\mathcal{R}$ into $I_\Delta \times J_\Delta \times K_\Delta$ uniform grids, each of size $\Delta \times \Delta \times \Delta$~(in meters), where $I_\Delta$, $J_\Delta$, and $K_\Delta$ denote the numbers of grids along the length, width, and height of $\mathcal{R}$, respectively, as illustrated in Fig.~\ref{Fig: 3D Environment}. 

When $\Delta$ is sufficiently small relative to the physical dimensions of the structures in $\mathcal{R}$, this discretization enables a compact yet faithful representation of the propagation environment. Specifically, the environment descriptor can be encoded as a binary tensor $\Bm^\Delta \in \{0,1\}^{I_\Delta \times J_\Delta \times K_\Delta}$, whose $(i,j,k)$-th is defined as
\begin{align} \label{Eq: Definition of B^Delta}
\Bm^\Delta(i,j,k) =
\begin{cases}
1, & \pb_{i,j,k} \in \mathcal{O}, \\
0, & \text{otherwise},
\end{cases}
\end{align}
where $\pb_{i,j,k}$ denotes the centroid of grid $(i,j,k)$. Similarly, the transmitter location is encoded as a one-hot tensor $\Qm^\Delta \in \{0,1\}^{I_\Delta \times J_\Delta \times K_\Delta}$, where $\Qm^\Delta(i,j,k) = 1$ if and only if $\qb_T$ lies within grid $(i,j,k)$.

Under this grid-based representation, the 3D RM over $\Rcal$ is represented by a real-valued tensor $\Pm_{\mathcal{R}}^{\Delta} \in \mathbb{R}^{I_{\Delta} \times J_{\Delta} \times K_{\Delta}}$, whose $(i,j,k)$-th entry is
\begin{align} \label{Eq: RM of Path Loss}
    \Pm_{\mathcal{R}}^{\Delta}(i,j,k)
    = P_L(\pb_{i,j,k}),
    \quad \forall\, i,j,k.
\end{align}
In this work, we adopt the path loss value at the centroid of each grid as the representative value for the entire grid under the assumption that the path loss varies negligibly within each grid~\cite{UAVPathPlan}.\footnote{In principle, one could adopt a
finer grid for the environment descriptor than for the 3D RM. However, this would substantially increase the data volume with marginal improvement in accuracy. We therefore use a common grid size for both $\Bm^{\Delta}$ and $\Pm^{\Delta}_{\mathcal{R}}$.} From~\eqref{Eq: PL function}, it follows that the 3D RM is fully determined by the discretized environment descriptor and the transmitter location:
\begin{align} \label{Eq: RM compact}
    \Pm_{\mathcal{R}}^{\Delta}
    = \mathcal{F}\left( \Bm^{\Delta}, \, \Qm^{\Delta} \right),
\end{align}
where $\mathcal{F}: ( \Bm^{\Delta},\,
\Qm^{\Delta} ) \mapsto \Pm_{\mathcal{R}}^{\Delta}$
denotes the propagation operator that aggregates the element-wise function $f(\cdot)$ across all grids.

With the above definitions, our objective is to learn a parameterized approximation of the propagation operator $\mathcal{F}(\cdot)$, such that, given the discretized environment descriptor $\Bm^{\Delta}$ and the transmitter location tensor $\Qm^{\Delta}$, the 3D RM $\Pm_{\mathcal{R}}^{\Delta}$ can be efficiently constructed.\footnote{Although we focus on path loss throughout this paper, the proposed framework readily extends to other channel quantities that are deterministic functions of the propagation environment and the transmitter location.}

\section{Practical Hybrid Dataset for 3D Environment} \label{Sec: Hybrid Dataset}
Recently, deep neural networks have shown great promise in learning complex propagation mappings owing to their powerful representation capabilities~\cite{RadioUNet, RMEGAN, RadioGAT, RadioDiff, RadioDiff3DSlice, SpectrumNet, SliceGNN, TianyuTVT, ZhiboLett, RadioDiff3D}. A dataset is necessary to train these models that can construct RMs with resolution $\Delta$ from the transmitter site and the propagation environment. One standard way is to build a dataset consisting of massive pairs of environment descriptors $\Bm^{\Delta}$, transmitter location maps $\Qm^{\Delta}$, and 3D RMs of resolution $\Delta$. However, in practice, it is too labor-intensive to measure path loss values at massive sites of each area to construct RMs in the dataset with the required resolution. 

In practice, obtaining 3D RMs at low resolutions is more feasible due to the substantially reduced number of measurement points for each area. Motivated by this observation, we propose to design a data-friendly DL architecture that is trained over a hybrid dataset consisting of RMs with heterogeneous resolutions. Specifically, let $\Delta$ denote the target high resolution and $\Delta_L$ denote a coarser resolution, where $\Delta_L > \Delta$. The practical hybrid dataset comprises the following two components:
\begin{itemize}
    \item \textbf{Low-resolution dataset $\mathcal{D}_L$}: RMs with resolution of $\Delta_L$ are constructed in all the areas, where channels are measured at coarsely spaced grids across all the areas.
    \item \textbf{High-resolution dataset $\mathcal{D}_H$}: RMs with the desirable resolution of $\Delta$ are merely present for a smaller number of areas, where channels are measured at densely spaced grids in certain areas.
\end{itemize}
To emulate such a practical hybrid dataset, this paper builds a simulated counterpart that preserves its main feature. The detailed construction procedure, including the RT setup, is described in Section~\ref{Sec: Dataset Construction}.

The technical challenge lies in how to leverage the above hybrid dataset $\mathcal{D} = \{\mathcal{D}_L, \mathcal{D}_H\}$ to learn the complex relationship among the environment descriptor $\Bm^{\Delta}$, the transmitter site $\Qm^{\Delta}$, and the high-resolution 3D RM $\Pm_{\Rcal}^{\Delta}$.

\section{Proposed Two-Stage DL Framework and Training Methodology} \label{Sec: Two-stage}
To effectively leverage the hybrid dataset, we propose a novel DL framework termed DF-3DRME for high-resolution 3D RM construction. The key idea of DF-3DRME is to decompose the challenging direct mapping problem into two complementary subproblems. The first stage focuses on capturing the fundamental radio propagation characteristics at a coarse spatial resolution, while the second stage aims to refine these coarse predictions by recovering the fine-grained spatial details.

Specifically, in the first stage, we design a compact neural network, termed LR-Net, to predict the low-resolution 3D RM, denoted as $\Pm_{\mathcal{R}}^{\Delta_L}$, from the environmental configuration $(\Bm^{\Delta}, \Qm^{\Delta})$. This stage leverages the abundant data samples in dataset $\mathcal{D}_L$ to learn the fundamental characteristics of radio propagation within the given environment. In the second stage, we employ a super-resolution network, termed SR-Net, to upscale the predicted low-resolution 3D RM to the high-resolution one at the target resolution of $\Delta$, i.e., $\Pm_{\mathcal{R}}^{\Delta}$. Notably, this stage requires substantially fewer high-fidelity training samples, as it focuses solely on refining the detailed characteristics captured by the low-resolution 3D RMs predicted by LR-Net rather than performing the direct environment-to-propagation mapping.

To further enhance the spatial fidelity of the predicted high-resolution 3D RMs, we carefully design the loss function that contains both of the reconstruction loss and perceptual loss in the training objective. Additionally, we further propose a multi-phase training strategy that combines stage-wise pretraining with end-to-end fine-tuning to maximize the learning efficiency and model performance. 

In the following subsections, we first detail the architectural design of both LR-Net and SR-Net, and then present the loss formulation along with the multi-phase training strategy for the proposed DF-3DRME.
\begin{figure*}[!t]
\centerline{\includegraphics[width=0.8\textwidth]{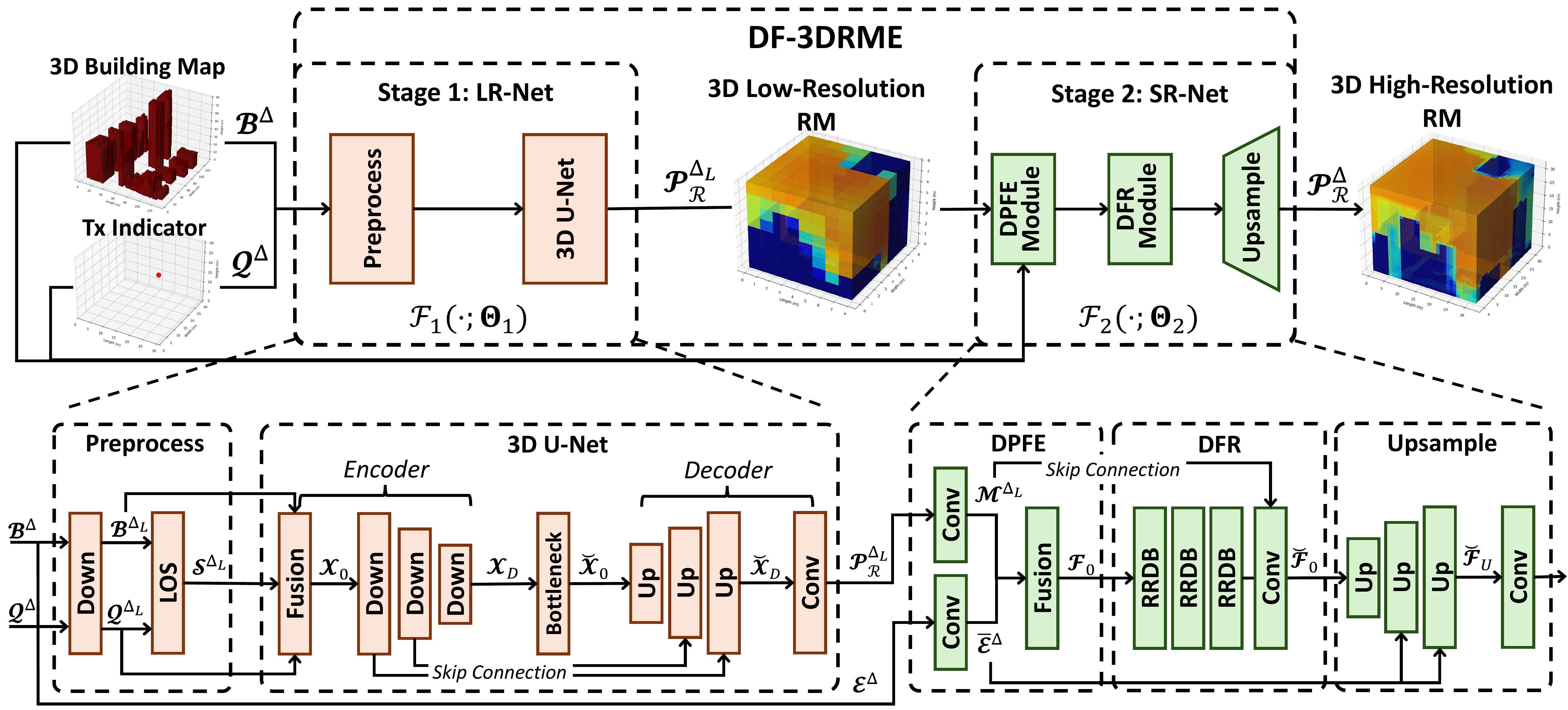}}
\vspace{-0.3cm}
\caption{Block diagram of DF-3DRME.}
\label{Fig: Block Diagram}
\vspace{-0.5cm}
\end{figure*}

\vspace{-0.3cm}
\subsection{Stage 1: Low-Resolution 3D RM Construction}
As illustrated in Fig.~\ref{Fig: Block Diagram}, LR-Net predicts the low-resolution 3D RM from $\Bm^{\Delta}$ and $\Qm^{\Delta}$ via two consecutive steps: i) data preprocessing, which reduces computational complexity and injects physics-informed priors, and ii) 3D U-Net processing, which maps the augmented input to the low-resolution 3D RM. The detailed operations are introduced as follows.

\subsubsection{Data Preprocessing}
Recall that the dimensions of both $\Bm^{\Delta}$ and $\Qm^{\Delta}$ are $I_{\Delta} \times J_{\Delta} \times K_{\Delta}$, which are considerably larger than those of the low-resolution 3D RM with dimension $I_{\Delta_L} \times J_{\Delta_L} \times K_{\Delta_L}$. To reduce memory consumption and improve computational efficiency, we first apply a deterministic downscaling operator $\mathcal{F}_1^{\Delta\rightarrow \Delta_L}(\cdot)$ that transforms tensors from resolution $\Delta$ to $\Delta_L$, which can be expressed as
\begin{align}
\Bm^{\Delta_L} &= \mathcal{F}_1^{\Delta\rightarrow \Delta_L}\left(\Bm^{\Delta}\right),\\
\Qm^{\Delta_L} &= \mathcal{F}_1^{\Delta\rightarrow\Delta_L}\left(\Qm^{\Delta}\right).
\end{align}
However, directly learning the mapping from the downscaled 3D EM and transmitter indicator $(\Bm^{\Delta_L},\Qm^{\Delta_L})$ to the low-resolution 3D RM $\Pm_{\mathcal{R}}^{\Delta_L}$ is still more challenging than its 2D counterpart, as 3D propagation depends on richer geometric information, including multi-floor blockage and height-dependent shadowing. To inject physically meaningful inductive bias into the model, we augment the input with an explicit LoS visibility tensor. Specifically, we define a binary LoS tensor $\Sm^{\Delta_L}\in\{0,1\}^{I_{\Delta_L}\times J_{\Delta_L}\times K_{\Delta_L}}$, where $\Sm^{\Delta_L}(i, j, k)=1$ if the discrete line segment from the transmitter to the centroid of the $(i,j,k)$-th grid is unobstructed by buildings, and $\Sm^{\Delta_L}(i, j, k)=0$ otherwise. This LoS tensor provides propagation-relevant visibility information, thereby reducing the learning complexity for the subsequent neural network.

To construct the LoS tensor, we employ a grid traversal algorithm based on the 3D Bresenham line algorithm~\cite{Bresenham}. Given the 3D EM $\Bm^{\Delta_L}$ and the transmitter location extracted from $\Qm^{\Delta_L}$, this algorithm determines whether the line segment connecting the transmitter and the centroid of each grid intersects any building grids, which can be expressed as
\begin{align}
\Sm^{\Delta_L} &= \mathcal{F}_1^{\rm LOS}\left(\Bm^{\Delta_L}, \, \Qm^{\Delta_L}\right),
\end{align}
where $\mathcal{F}_1^{\rm LOS}(\cdot)$ denotes the 3D Bresenham-based visibility extraction operator. Notably, this operation can be executed efficiently due to the reduced spatial dimensions of $\Bm^{\Delta_L}$ and $\Qm^{\Delta_L}$ after downscaling.

\subsubsection{3D U-Net Processing}
The U-Net architecture is well-suited for 3D RM prediction for two reasons. First, its hierarchical structure enables the model to capture multi-scale propagation phenomena from local shadowing effects to large-scale path loss patterns, which is essential for accurate 3D RM prediction. Second, the U-Net architecture has demonstrated remarkable effectiveness in 2D RM estimation~\cite{RadioUNet}, thereby motivating a natural extension to the 3D scenario.

Given the input tensors $\Bm^{\Delta_L}$, $\Qm^{\Delta_L}$, and $\Sm^{\Delta_L}$, we adopt a 3D U-Net architecture as the backbone for learning the mapping from $(\Bm^{\Delta_L}, \Qm^{\Delta_L}, \Sm^{\Delta_L})$ to $\Pm_{\mathcal{R}}^{\Delta_L}$. As shown in Fig.~\ref{Fig: Block Diagram}, the proposed 3D U-Net follows an encoder-bottleneck-decoder structure with specialized attention mechanisms and residual connections to enhance feature learning capability and gradient flow stability. In the following, we introduce each component in detail.

\textbf{Encoder Path:} 
The inputs $\Bm^{\Delta_L}$, $\Qm^{\Delta_L}$, and $\Sm^{\Delta_L}$ are first individually processed through separate convolutional blocks to extract domain-specific features, which are then fused to form an initial feature tensor $\Xm_0$. Subsequently, the encoder employs $D$ downsampling blocks to progressively compress $\Xm_0$ into higher-level representations, each consisting of a 3D residual attention block followed by a max pooling layer. More specifically, each residual attention block integrates a 3D convolutional block attention module (CBAM)~\cite{CBAM} that applies sequential channel and spatial attention to focus selectively on propagation-relevant features, while residual connections stabilize gradient flow~\cite{ResNet}. Let $\Xm_d$, $\forall d \in [D]$, denote the output feature tensor of the $d$-th downsampling block. The encoder operation can be summarized as
\begin{align}
    \Xm_{D} = \mathcal{F}_1^{\rm ENC} \left( \Bm^{\Delta_L}, \,\Qm^{\Delta_L}, \, \Sm^{\Delta_L}; \mathbf{\Theta}_1^{\rm ENC} \right),
\end{align}
where $\Xm_{D}$ denotes the output feature tensor after $D$ downsampling blocks, and $\mathcal{F}_1^{\rm ENC} \left( \cdot; \mathbf{\Theta}_1^{\rm ENC} \right)$ represents the encoder function parameterized by $\mathbf{\Theta}_1^{\rm ENC}$. The feature tensor $\Xm_{D}$ is then passed to a bottleneck module for further abstraction and global context aggregation.

\textbf{Bottleneck Module:}
The bottleneck module operates on the compressed feature representation and serves to capture global contextual dependencies that span the entire spatial domain. This is particularly important for radio propagation modeling, where signal characteristics at one location can be influenced by distant environmental structures. The bottleneck module consists of three key components: i) a stack of 3D residual blocks for deep feature transformation, ii) a self-attention mechanism for capturing long-range spatial dependencies across the entire 3D space, and iii) a dropout layer for regularization to mitigate overfitting and improve generalization. Finally, a shallow 3D convolutional block further refines the feature tensor. Let $\breve{\Xm}_0$ denote the output feature tensor of the bottleneck module. The overall bottleneck processing can be summarized as
\begin{align}
    \breve{\Xm}_0 = \mathcal{F}_1^{\rm BOT} \left( \Xm_D; \,\mathbf{\Theta}_1^{\rm BOT} \right),
\end{align}
where $\mathcal{F}_1^{\rm BOT} \left( \cdot; \mathbf{\Theta}_1^{\rm BOT} \right)$ denotes the composite bottleneck function and $\mathbf{\Theta}_1^{\rm BOT}$ contains all parameters of the bottleneck module. The refined feature tensor $\breve{\Xm}_0$ is subsequently fed to the decoder for reconstructing the low-resolution 3D RM.

\textbf{Decoder Path:} 
The decoder reconstructs the low-resolution 3D RM $\Pm_{\mathcal{R}}^{\Delta_L}$ from the bottleneck feature tensor $\breve{\Xm}_0$ through $D$ upsampling blocks. Each upsampling block consists of three key components: i) a transposed convolution to increase the spatial dimension, ii) a skip connection that concatenates feature tensors from the corresponding encoder stage to preserve fine-grained spatial details, and iii) a residual attention block that fuses the concatenated features. After $D$ upsampling blocks, an output refinement module with 3D CBAM maps the final feature tensor $\breve{\Xm}_{D}$ to a single-channel tensor, which is then reshaped to generate the low-resolution 3D RM. The overall decoder operation can be formulated as
\begin{align}
    \breve{\Xm}_{D} = \mathcal{F}_1^{\rm DEC} \left( \breve{\Xm}_0, \, \left\{ \Xm_n \right\}_{n=1}^{D};\, \mathbf{\Theta}_1^{\rm DEC} \right),
\end{align}
where $\mathcal{F}_1^{\rm DEC}(\cdot; \mathbf{\Theta}_1^{\rm DEC})$ denotes the decoder function parameterized by $\mathbf{\Theta}_1^{\rm DEC}$.

By composing the aforementioned operations, the overall processing pipeline of LR-Net can be summarized as
\begin{align}
\Pm_{\mathcal{R}}^{\Delta_L} = \mathcal{F}_{1}(\Bm^{\Delta}, \, \Qm^{\Delta} \,; \, \boldsymbol{\Theta}_1),
\end{align}
where $\mathcal{F}_1(\cdot; \boldsymbol{\Theta}_1)$ denotes the composite LR-Net function defined as
\begin{align}
\mathcal{F}_1 \triangleq \mathcal{F}_1^{\rm DEC} \circ \mathcal{F}_1^{\rm BOT} \circ \mathcal{F}_1^{\rm ENC} \circ \left( \mathcal{F}_1^{\Delta \rightarrow \Delta_L},\, \mathcal{F}_1^{\rm LOS} \right),
\end{align}
and $\mathbf{\Theta}_1 = \{ \mathbf{\Theta}_1^{\rm ENC}, \mathbf{\Theta}_1^{\rm BOT}, \mathbf{\Theta}_1^{\rm DEC} \}$ encompasses all trainable parameters in LR-Net. By leveraging abundant low-resolution training samples in the hybrid dataset, LR-Net learns to produce accurate low-resolution 3D RMs $\Pm_{\mathcal{R}}^{\Delta_L}$, which serve as priors for the subsequent super-resolution stage. 

\vspace{-0.3cm}
\subsection{Stage~2: High-Resolution 3D RM Construction via Super Resolution}
The second stage employs a 3D super-resolution network to reconstruct the high-resolution 3D RM $\Pm_{\mathcal{R}}^{\Delta}$ from the low-resolution counterpart $\Pm_{\mathcal{R}}^{\Delta_L}$ produced by LR-Net, along with the environment configuration $(\Bm^{\Delta}, \Qm^{\Delta})$. The design of SR-Net is motivated by two key priors: i) the low-resolution 3D RM $\Pm_{\mathcal{R}}^{\Delta_L}$ already captures the dominant propagation effects and thus provides a reliable initialization for high-resolution 3D RM reconstruction, and ii) the building layout and transmitter location offer fine-grained structural constraints that guide the recovery of spatial details in the high-resolution 3D RM. To effectively exploit these priors, we implement the SR-Net with a three-module architecture:
\begin{enumerate}
    \item \textbf{Dual-Path Feature Extraction (DPFE) Module:} This module employs two parallel pathways to independently extract features from the low-resolution 3D RM $\Pm_{\mathcal{R}}^{\Delta_L}$ and the environment configuration $(\Bm^{\Delta}, \Qm^{\Delta})$, followed by an attention-enhanced fusion mechanism that integrates these heterogeneous representations.
    \item \textbf{Dense Feature Refinement (DFR) Module:} This module leverages a cascade of 3D Residual-in-Residual Dense Blocks (RRDBs)~\cite{ESRGAN} with multi-level residual learning and dense connections to progressively enhance the feature representation capacity.
    \item \textbf{HR Generation Module:} This module reconstructs the high-resolution 3D RM through progressive upsampling operations while incorporating multi-scale environmental features to ensure geometric consistency at each stage.
\end{enumerate}
In the following, we introduce the implementation of each module in detail.

\subsubsection{DPFE Module}
A key observation is that different input modalities contribute distinct information to the super-resolution task: the low-resolution 3D RM $\Pm_{\mathcal{R}}^{\Delta_L}$ provides the fundamental radio propagation characteristics, whereas the environmental configuration $(\Bm^{\Delta}, \Qm^{\Delta})$ provide fine-grained geometric context. Moreover, unlike LR-Net which requires an explicit LoS feature tensor to establish transceiver visibility relationships, SR-Net can omit this auxiliary input since the low-resolution 3D RM already encodes LoS information through the learned propagation patterns. To effectively leverage these heterogeneous inputs, the DPFE module adopts a dual-path architecture that processes each input modality through specialized feature extractors.

First, we concatenate $\Bm^{\Delta}$ and $\Qm^{\Delta}$ along the channel axis to form the environmental input $\Em^{\Delta} \in \mathbb{R}^{2 \times I_{\Delta} \times J_{\Delta} \times K_{\Delta}}$. The DPFE module then employs two complementary pathways to process the low-resolution 3D RM input $\Pm_{\mathcal{R}}^{\Delta_L}$ and the environmental input $\Em^{\Delta}$ separately. Specifically, the RM pathway processes $\Pm_{\mathcal{R}}^{\Delta_L}$ through a single 3D convolutional layer to extract the radio feature, denoted by $\Mm^{\Delta_L} \in \mathbb{R}^{C \times I_{\Delta_L} \times J_{\Delta_L} \times K_{\Delta_L}}$, where $C$ denotes the number of base channels. In parallel, the environment pathway processes $\Em^{\Delta}$ through a sequence of 3D convolutional layers to produce the environmental feature, denoted by $\overline{\Em}^{\Delta} \in \mathbb{R}^{C \times I_{\Delta} \times J_{\Delta} \times K_{\Delta}}$.

To fuse the radio feature $\Mm^{\Delta_L}$ with the environmental feature $\overline{\Em}^{\Delta}$, we must first address the spatial dimension mismatch between them. To this end, we employ the adaptive average pooling operation, which dynamically computes the pooling kernel size and stride, to downscale $\overline{\Em}^{\Delta}$ to have the same spatial dimension as $\Mm^{\Delta_L}$:
\begin{align}
\overline{\Em}^{\Delta_L} = \Pcal\left(\overline{\Em}^{\Delta}; \, \Mm^{\Delta_L}\right),
\end{align}
where $\Pcal(\cdot; \triangleright)$ is the adaptive average pooling function that automatically downscales the first input tensor to match the spatial dimension of the second one, and $\overline{\Em}^{\Delta_L} \in \mathbb{R}^{C \times I_{\Delta_L} \times J_{\Delta_L} \times K_{\Delta_L}}$ is the downscaled environmental feature.

Given the dimension-aligned radio feature $\Mm^{\Delta_L}$ and environmental feature $\overline{\Em}^{\Delta_L}$, we employ a fusion module to integrate them. The fusion module first concatenates $\Mm^{\Delta_L}$ and $\overline{\Em}^{\Delta_L}$ along the channel dimension, and then applies a 3D convolutional layer to reduce the channel dimension back to $C$. The fused feature is subsequently refined by a 3D CBAM. Let $\Fm_0$ denote the feature tensor after this fusion process, which also serves as the output of the DPFE module. Thus, the entire DPFE process can be formulated as
\begin{align}
    \Fm_0 = \mathcal{F}_2^{\rm DPFE}\left( \Bm^{\Delta}, \, \Qm^{\Delta}, \, \Pm_{\mathcal{R}}^{\Delta_L}; \, \mathbf{\Theta}_2^{\rm DPFE} \right),
\end{align}
where $\mathbf{\Theta}_2^{\rm DPFE}$ stores all the trainable parameters in the DPFE module. The feature tensor $\Fm_0$ is then fed to the DFR module for enhanced representation learning.

\subsubsection{DFR Module}
The function of the DFR module is to enhance the representation capacity of the fused features through deep feature extraction. To achieve this goal, we adopt the RRDB architecture~\cite{ESRGAN}, which offers several advantages: i) dense connections within each Residual Dense Block (RDB) facilitate feature reuse, ii) multi-level residual learning stabilizes training of deep networks, and iii) residual scaling prevents gradient explosion while maintaining network expressiveness. We describe the pipeline of the $g$-th RRDB, $\forall g \in [G]$, as an example to elucidate the mechanism of the DFR module.

\emph{Example:} Without loss of generality, let $\Fm_{g-1}$ denote the feature tensor after the $(g-1)$-th RRDB, which serves as the input to the $g$-th RRDB. Each RRDB consists of three cascaded RDBs, where each RDB is implemented as a stack of 3D convolutional layers with dense connections. Specifically, within each RDB, every convolutional layer receives the concatenation of all preceding feature maps as input, and its output is passed to all subsequent layers, thereby facilitating extensive feature reuse. The outputs from all dense layers are aggregated through a local feature fusion layer, and a local residual connection is added to preserve the input information. After passing through the three RDBs, a residual scaling factor of $0.2$ is applied to the output before adding the global skip connection. Therefore, the process of the $g$-th RRDB can be compactly summarized as
\begin{align} \label{Eqn: F_g}
\Fm_g = 0.2 \cdot \mathrm{RRDB}_g(\Fm_{g-1}) + \Fm_{g-1}, \quad \forall g,
\end{align}
where $\mathrm{RRDB}_g(\cdot)$ represents the composite function in the $g$-th RRDB, and $\Fm_g$ is the output feature tensor. Note that the residual scaling factor of $0.2$ is empirically chosen to stabilize training and prevent gradient explosion in deep networks~\cite{ESRGAN}. For notational simplicity, we omit the dependence on the trainable parameters for \eqref{Eqn: F_g}, which differ across the RRDBs.

After processing through $G$ RRDBs, we obtain $\Fm_G$ as the deeply extracted feature tensor. Subsequently, we apply an additional 3D convolutional layer for feature aggregation and introduce a long-range global residual connection from the radio feature $\Mm^{\Delta_L}$. This global residual connection serves to preserve the low-frequency content encoded in the original coarse RM~\cite{ESRGAN}. The aggregated feature can be expressed as
\begin{align}
\breve{\Fm}_0 = {\rm Conv3D}\left(\Fm_G\right) + \Mm^{\Delta_L},
\end{align}
where $\breve{\Fm}_0$ denotes the deeply refined feature tensor. Note that the trainable parameters in this 3D convolutional block differs from that of other modules and is omitted for notational simplicity. Therefore, the overall DFR module can be compactly summarized as
\begin{align}
\breve{\Fm}_0 = \mathcal{F}_2^{\rm DFR} \left( \Fm_0, \, \Mm^{\Delta_L}; \, \mathbf{\Theta}_2^{\rm DFR} \right),
\end{align}
where $\mathbf{\Theta}_2^{\rm DFR}$ stores all the trainable parameters in the DFR module, including $G$ RRDBs and the final convolutional layer. The refined feature tensor $\breve{\Fm}_0$ is then fed to the HR generation module for spatial resolution enhancement.

\subsubsection{HR Generation Module}
The HR generation module is responsible for progressively reconstructing the high-resolution 3D RM from the deeply refined feature $\breve{\Fm}_0$. A critical design consideration is the choice of upsampling strategy. Conventional approaches such as trilinear interpolation usually generate overly smooth outputs~\cite{SubPixel}. To address these issues, we adopt \emph{voxel shuffle}, a 3D extension of the sub-pixel convolution~\cite{SubPixel}. Given a feature tensor, the voxel shuffle operation first expands the channel dimension through convolution, then rearranges channel elements into spatial entries, allowing efficient learning of optimal upsampling filters from data.

Based on the refined feature tensor $\breve{\Fm}_0$, we reconstruct the high-resolution 3D RM through $U$ consecutive upsampling blocks. Each upsampling block comprises two key components: i) a voxel shuffle upsampling operator that increases the spatial dimensions by a factor of $2$, and ii) a fusion block that integrates environmental features $\overline{\Em}^{\Delta}$ at the corresponding spatial scale. We illustrate this process by using the first upsampling block as an example.

\emph{Example}: Starting from the refined feature tensor $\breve{\Fm}_0 \in \mathbb{R}^{C\times I_{\Delta_L} \times J_{\Delta_L} \times K_{\Delta_L}}$, we first apply a 3D convolution that expands the channel dimension from $C$ to $8C$. Subsequently, the voxel shuffle operation rearranges these $8C$ channels into a $2\times2\times2$ spatial grid per output voxel to double each spatial dimension while reducing the channel count back to $C$. Let $\Xm_1 \in \mathbb{R}^{C \times 2I_{\Delta_L} \times 2J_{\Delta_L} \times 2K_{\Delta_L}}$ denote the output tensor of the voxel shuffle operation in the first upsampling stage. After the upsampling operation, we fuse $\Xm_1$ with the environmental features $\overline{\Em}^{\Delta}$. By incorporating environmental features at each upsampling block, we provide explicit geometric guidance that ensures the reconstructed RM adheres to the physical structure of the propagation environment. This step is essential because as spatial resolution increases, the network must recover increasingly fine geometric details such as building edges and corner diffraction effects, which cannot be accurately inferred from the low-resolution RM alone. Since the spatial dimension of $\overline{\Em}^{\Delta}$ may be larger than that of $\Xm_1$ at intermediate upsampling blocks, we again employ adaptive average pooling to align their dimensions
\begin{align}
\overline{\Em}_1 = \mathcal{P} \left( \overline{\Em}^{\Delta}; \, \Xm_1 \right),
\end{align}
where $\overline{\Em}_1$ is the rescaled environmental feature matching the spatial dimension of $\Xm_1$. Subsequently, $\Xm_1$ and $\overline{\Em}_1$ are concatenated along the channel axis and processed through a fusion module consisting of a 3D convolutional layer followed by a 3D CBAM block. This produces the fused feature $\breve{\Fm}_1 \in \mathbb{R}^{C\times 2I_{\Delta_L} \times 2J_{\Delta_L} \times 2K_{\Delta_L}}$, which serves as the output of the first upsampling block.

By repeating these operations for $U$ upsampling blocks, we obtain the final feature tensor $\breve{\Fm}_U \in \mathbb{R}^{C \times I_{\Delta} \times J_{\Delta} \times K_{\Delta}}$. Finally, we employ a refinement sub-network containing a stack of 3D convolutional layers to generate the 3D RM at target resolution $\Delta$, denoted as $\widetilde{\Pm}_{\mathcal{R}}^{\Delta}$. To further enhance the robustness, we introduce a learnable scalar $\alpha$ for output calibration, which rescales the reconstructed RM as $\Pm_{\mathcal{R}}^{\Delta} = \alpha \cdot \widetilde{\Pm}_{\mathcal{R}}^{\Delta}$. Empirically, this learnable scaling factor is particularly beneficial for preventing suboptimal magnitude estimation when only a limited amount of training data is available. Thus, the overall HR generation module can be expressed as
\begin{align}
\Pm_{\mathcal{R}}^{\Delta} = \mathcal{F}_2^{\rm HR} \left( \breve{\Fm}_0, \, \left\{ \overline{\Em}_u \right\}_{u=1}^{U}; \, \mathbf{\Theta}_2^{\rm HR} \right),
\end{align}
where $\mathbf{\Theta}_2^{\rm HR}$ collects all trainable parameters in the HR generation module. 

Combining all three modules, the overall super-resolution process in SR-Net can be compactly expressed as
\begin{align}
\Pm_{\mathcal{R}}^{\Delta}
= \mathcal{F}_2\left(\Bm^{\Delta}, \, \Qm^{\Delta}, \, \Pm_{\mathcal{R}}^{\Delta_L}; \, \mathbf{\Theta}_2\right),
\end{align}
where $\mathcal{F}_2$ denotes the composite mapping given by
\begin{align}
\mathcal{F}_2 \triangleq \mathcal{F}_2^{\rm HR} \circ \mathcal{F}_2^{\rm DFR} \circ \mathcal{F}_2^{\rm DPFE},
\end{align}
and $\mathbf{\Theta}_2 = \{\mathbf{\Theta}_2^{\rm DPFE}, \mathbf{\Theta}_2^{\rm DFR}, \mathbf{\Theta}_2^{\rm HR}\}$ collects all learnable parameters from the DPFE, DFR, and HR generation modules.

\vspace{-0.4cm}
\subsection{Training Policy}
The training policy is essential for the high-resolution 3D RM construction performance of the proposed DF-3DRME. In this work, the training policy comprises two components: i) a composite loss function that balances voxel-wise reconstruction accuracy with perceptual quality, and ii) a multi-phase training strategy that ensures effective learning at each stage while maintaining end-to-end compatibility. In the following, we elaborate on each component in detail.

\subsubsection{Loss Function Design}
We employ a composite loss function to optimize both voxel-wise accuracy and perceptual quality of the predicted 3D RMs. Specifically, for a 3D RM with spatial resolution $\Delta$, the total loss function is defined as
\begin{align} \label{Eq: Combined Loss}
\mathcal{L}^{\Delta} = \mathcal{L}^{\Delta}_{\rm MSE} + \lambda \mathcal{L}^{\Delta}_{\ell_1} + \gamma \mathcal{L}^{\Delta}_{\rm Percep.},
\end{align}
where $\mathcal{L}^{\Delta}_{\rm MSE}$, $\mathcal{L}^{\Delta}_{\ell_1}$, and $\mathcal{L}^{\Delta}_{\rm Percep.}$ denote the mean squared error (MSE) loss, the $\ell_1$ loss, and the perceptual loss, respectively, while $\lambda$ and $\gamma$ are hyperparameters that balance the contributions of each loss component. In what follows, we describe each loss component in detail.

\textbf{MSE Loss:}
The MSE loss measures the average squared difference between the predicted and ground truth 3D RMs, and is defined as
\begin{align}
\mathcal{L}^{\Delta}_{\rm MSE} = \mathbb{E} \left[ \dfrac{1}{I_{\Delta}J_{\Delta}K_{\Delta}} \left\| \widehat{\Pm}_{\mathcal{R}}^{\Delta} - \Pm_{\mathcal{R}}^{\Delta} \right\|_F^2 \right],
\end{align}
where $\widehat{\Pm}_{\mathcal{R}}^{\Delta}$ and $\Pm_{\mathcal{R}}^{\Delta}$ denote the predicted and ground truth 3D RMs for $\mathcal{R}$, respectively, and the expectation is approximated by averaging over all training samples.

\textbf{$\ell_1$ Loss:}
The $\ell_1$ loss computes the average absolute difference between the predicted and ground truth 3D RMs, which is given by
\begin{align}
\mathcal{L}^{\Delta}_{\ell_1} = \mathbb{E} \left[ \dfrac{1}{I_{\Delta} J_{\Delta} K_{\Delta}} \left\| \widehat{\Pm}_{\mathcal{R}}^{\Delta} - \Pm_{\mathcal{R}}^{\Delta} \right\|_1 \right].
\end{align}
Compared to the MSE loss, the $\ell_1$ loss is less sensitive to outliers and helps preserve edge sharpness in the predicted 3D RMs~\cite{Perceptual}.

\textbf{Perceptual Loss:}
The perceptual loss is incorporated to capture high-level feature similarities between the predicted and ground truth 3D RMs, thereby enhancing the visual quality and structural fidelity of the reconstructed RMs~\cite{Perceptual}. Since existing perceptual loss formulations rely on feature extractors pretrained on 2D natural images (e.g., VGG networks~\cite{VGG}), we propose an altitude-averaged perceptual loss tailored for 3D RMs. Specifically, we decompose the 3D RM into a sequence of 2D horizontal slices along the altitude dimension. For each altitude index $k \in [K_{\Delta}]$, we extract the corresponding 2D slice $\widehat{\Pb}_{\mathcal{R},k}^{\Delta}= \widehat{\Pm}_{\mathcal{R}}^{\Delta}(:,:,k) \in \mathbb{R}^{I_{\Delta} \times J_{\Delta}}$ from the predicted RM and $\Pb_{\mathcal{R},k}^{\Delta} = \Pm_{\mathcal{R}}^{\Delta}(:,:,k) \in \mathbb{R}^{I_{\Delta} \times J_{\Delta}}$ from the ground truth. Let $\phi_l(\cdot)$ denote the feature map extracted from the $l$-th layer of a pretrained VGG-16 network~\cite{VGG}. The perceptual loss for each 2D slice is computed as the MSE between the feature representations, and the overall perceptual loss is obtained by averaging across all altitude slices and all selected VGG layers:
\begin{align}
\mathcal{L}^{\Delta}_{\rm Percep.} = \mathbb{E} \left[ \dfrac{1}{K_{\Delta}L} \sum_{k=1}^{K_{\Delta}} \sum_{l=1}^{L} \left\| \phi_l \left( \widehat{\Pb}_{\mathcal{R},k}^{\Delta} \right) - \phi_l\left( \Pb_{\mathcal{R},k}^{\Delta} \right) \right\|_F^2 \right],
\end{align}
where $L$ denotes the number of selected VGG layers.

\subsubsection{Multi-Phase Training Strategy}
Given the cascaded architecture of our two-stage model, the end-to-end training can easily lead to suboptimal convergence performance. To address this issue, we adopt a multi-phase training strategy that progressively trains the neural networks in each stage. The training procedure consists of three phases. In the first phase, we pretrain LR-Net in Stage~1, i.e., $\mathcal{F}_1(\cdot; \mathbf{\Theta}_1)$, using loss $\mathcal{L}^{\Delta_L}$ with abundant low-resolution 3D RMs. This phase establishes the coarse prediction that serves as the robust priors for the subsequent super-resolution phase. In the second phase, we pretrain the SR-Net in Stage~2, i.e., $\mathcal{F}_2(\cdot; \mathbf{\Theta}_2)$, using loss $\mathcal{L}^{\Delta}$ with \emph{ground truth} low-resolution 3D RMs as inputs. This strategy allows Stage~2 to learn the optimal super-resolution mapping under ideal conditions. In the third phase, the SR-Net in Stage~2 is fine-tuned with the predicted low-resolution 3D RMs from the pretrained LR-Net as inputs, while keeping LR-Net's parameters frozen. This phase enhances the robustness of SR-Net in Stage~2 to the actual distribution of Stage~1's outputs, thus ensuring effective end-to-end performance. The detailed steps of this training approach are outlined in Algorithm \ref{Alg: Training Strategy}.
\vspace{-0.2cm}
\begin{algorithm}[htbp!]
\caption{Multi-Phase Training Strategy}
\label{Alg: Training Strategy}
\begin{algorithmic}[1]
\REQUIRE Low-resolution dataset $\mathcal{D}_L$; High-resolution dataset $\mathcal{D}_H$; Number of epochs $T_1$, $T_2$, $T_3$ for each phase
\ENSURE Trained parameters $\mathbf{\Theta}_1^*$, $\mathbf{\Theta}_2^*$
\STATE \textbf{// Phase 1: Pretrain LR-Net in Stage~1}
\STATE Initialize $\mathbf{\Theta}_1$ randomly
\FOR{epoch $= 1$ to $T_1$}
    \FOR{each mini-batch from $\mathcal{D}_L$}
        \STATE $\widehat{\Pm}_{\Rcal}^{\Delta_L} \leftarrow \mathcal{F}_1(\Bm^{\Delta}, \, \Qm^{\Delta}; \, \mathbf{\Theta}_1)$
        \STATE Compute loss $\mathcal{L}^{\Delta_L}(\widehat{\Pm}_{\Rcal}^{\Delta_L}, \, \Pm_{\Rcal}^{\Delta_L})$
        \STATE Update $\mathbf{\Theta}_1$ via gradient descent
    \ENDFOR
\ENDFOR
\STATE $\mathbf{\Theta}_1^* \leftarrow \mathbf{\Theta}_1$ \hfill \COMMENT{Save pretrained LR-Net}
\STATE
\STATE \textbf{// Phase 2: Pretrain SR-Net in Stage~2}
\STATE Initialize $\mathbf{\Theta}_2$ randomly
\FOR{epoch $= 1$ to $T_2$}
    \FOR{each mini-batch from $\mathcal{D}_H$}
        \STATE $\widehat{\Pm}_{\Rcal}^{\Delta} \leftarrow \mathcal{F}_2(\Bm^{\Delta}, \, \Qm^{\Delta}, \, \Pm_{\Rcal}^{\Delta_L}; \, \mathbf{\Theta}_2)$
        \STATE Compute loss $\mathcal{L}^{\Delta}(\widehat{\Pm}_{\Rcal}^{\Delta}, \, \Pm_{\Rcal}^{\Delta})$
        \STATE Update $\mathbf{\Theta}_2$ via gradient descent
    \ENDFOR
\ENDFOR
\STATE
\STATE \textbf{// Phase 3: Fine-tune SR-Net in Stage~2}
\FOR{epoch $= 1$ to $T_3$}
    \FOR{each mini-batch from $\mathcal{D}_H$}
        \STATE $\widehat{\Pm}_{\Rcal}^{\Delta_L} \leftarrow \mathcal{F}_1(\Bm^{\Delta}, \, \Qm^{\Delta}; \, \mathbf{\Theta}_1^*)$ \hfill \COMMENT{Frozen LR-Net}
        \STATE $\widehat{\Pm}_{\Rcal}^{\Delta} \leftarrow \mathcal{F}_2(\Bm^{\Delta}, \, \Qm^{\Delta}, \, \widehat{\Pm}_{\Rcal}^{\Delta_L}; \, \mathbf{\Theta}_2)$ 
        \STATE Compute loss $\mathcal{L}^{\Delta}(\widehat{\Pm}_{\Rcal}^{\Delta}, \, \Pm_{\Rcal}^{\Delta})$
        \STATE Update $\mathbf{\Theta}_2$ via gradient descent
    \ENDFOR
\ENDFOR
\STATE $\mathbf{\Theta}_2^* \leftarrow \mathbf{\Theta}_2$ \hfill \COMMENT{Save fine-tuned SR-Net}
\RETURN $\mathbf{\Theta}_1^*$, $\mathbf{\Theta}_2^*$
\end{algorithmic}
\end{algorithm}
\vspace{-0.5cm}
\begin{remark}
It is worth noting that the proposed DF-3DRME is agnostic to specific network architectures and can accommodate various backbone designs for both LR-Net and SR-Net. For example, in this work, we adopt a U-Net-based backbone for the LR-Net in Stage~1, which has been demonstrated to be effective for RM construction tasks~\cite{RadioUNet}. Nevertheless, the LR-Net can be readily replaced with more advanced architectures, such as vision transformers~\cite{ViT} or diffusion models~\cite{Diffusion}, to potentially achieve further performance improvements. The investigation of different backbone architectures is beyond the scope of this work and is left for future research.
\end{remark}

\vspace{-0.4cm}
\section{Experimental Results} \label{Sec: Experimental Results} \label{Sec: Experiments}
\subsection{Dataset Construction} \label{Sec: Dataset Construction}
We construct a hybrid 3D RM dataset comprising $N = 517$ distinct urban environments, each associated with $T = 100$ randomly deployed transmitter sites, resulting in a total of $NT = 51{,}700$ environment–transmitter pairs. The 3D EMs are derived from building footprints extracted from \textit{OpenStreetMap} (OSM)~\cite{OSM}, covering metropolitan areas including Berlin, Paris, Los Angeles, and New York. Each environment spans a cubic volume of $128~\text{m} \times 128~\text{m} \times 128~\text{m}$. The building heights are sampled independently and uniformly from the interval $[40, 100]~\text{m}$, while the transmitter heights are drawn uniformly from $[30, 80]~\text{m}$ to accommodate both terrestrial and aerial base station deployment scenarios.

The hybrid dataset consists of two components: a large-scale low-resolution dataset $\mathcal{D}_L$ and a smaller high-resolution dataset $\mathcal{D}_H$. To construct the low-resolution dataset $\mathcal{D}_L$, we set the resolution as $\Delta_L=4$ (in meters). Consequently, the spatial dimension of the low-resolution 3D RM for each urban environment is $I_{\Delta_L} = J_{\Delta_L} = K_{\Delta_L} = 32$. For each environment and each of its associated transmitters, we place a single-antenna receiver at the centroid of every grid and run the RT simulation to obtain the corresponding path loss value. This procedure yields $NT=51{,}700$ low-resolution 3D RMs. For the high-resolution dataset $\mathcal{D}_H$, we randomly select $M=60$ environments and set the resolution as $\Delta=1$, resulting in RMs of spatial dimension $I_{\Delta} = J_{\Delta} = K_{\Delta} = 128$. The same RT simulation procedure is applied to generate $MT=6{,}000$ high-resolution 3D RMs. By design, the high-resolution dataset is substantially smaller than the low-resolution counterpart, reflecting the practical constraint of limited availability of high-resolution 3D RMs.

All RT simulations are conducted using the commercial software WinProp~\cite{WinProp} with the dominant path model, which retains the most significant propagation paths and thus offers a favorable trade-off between computational efficiency and prediction accuracy. The simulation parameters are configured as follows: carrier frequency of $3.5~\text{GHz}$, system bandwidth of $10~\text{MHz}$, and transmit power of $23~\text{dBm}$. Both transmitter and receiver antennas are modeled as isotropic, which aligns with prior works~\cite{RadioUNet, RadioDiff3D}. To facilitate neural network training and ensure numerical stability, each 3D RM is normalized entry-wise to the range $[0, 1]$ following the procedures in~\cite{RadioUNet}.

The dataset is partitioned to support the proposed multi-phase training strategy. For Stage~1 pretraining, the low-resolution dataset $\mathcal{D}_L$ is randomly split into $437$ environments for training and $80$ for validation. For Stage~2 pretraining and fine-tuning, the high-resolution dataset $\mathcal{D}_H$ is divided into $40$ environments for training and $20$ for validation. For the testing phase, we construct a separate test set consisting of $80$ additional urban environments, each with $T = 100$ transmitter sites. Both low-resolution and high-resolution 3D RMs are generated for this test set via RT simulation. Note that the environments in the test set are non-overlapping with those in the training and validation sets to ensure a rigorous evaluation of generalization to unseen building geometries and transmitter configurations.

\vspace{-0.4cm}
\subsection{Setup for DL Model and Training Process}
The proposed DF-3DRME is implemented in PyTorch and executed on a workstation equipped with an AMD Ryzen 5795WX CPU and an NVIDIA RTX 5090 GPU. Both stages are trained using the AdamW optimizer with an adaptive learning rate schedule that combines linear warmup with plateau-based reduction to promote stable convergence. The weighting coefficients in the combined loss function~\eqref{Eq: Combined Loss} are set to $\lambda = 1$ and $\gamma = 0.2$.

Following the multi-phase training strategy, the training process is configured as follows. In the first two phases of pretraining Stage~1 and Stage~2, the initial learning rate is set to $2 \times 10^{-4}$. In the third phase, which fine-tunes Stage~2 in an end-to-end manner, the initial learning rate is reduced to $1 \times 10^{-5}$ to preserve the learned representations while allowing gradual adaptation to the cascaded pipeline. The batch size is set to $32$ for Stage~1 pretraining, which is feasible due to the reduced memory footprint of low-resolution 3D RM. For Stage~2 pretraining and fine-tuning, the batch size is reduced to $4$ to accommodate the larger memory requirements of high-resolution 3D RMs. Complete implementation details and reproducible code are at the project repository.\footnote{https://github.com/LINZHU-PolyU/DF-3DRME}

\vspace{-0.25cm}
\subsection{Performance Metrics}
We evaluate the proposed method using four standard metrics commonly adopted in prior RM reconstruction studies: normalized MSE (NMSE), root MSE (RMSE), structural similarity index measure (SSIM), and peak signal-to-noise ratio (PSNR). These metrics collectively assess the reconstruction accuracy from complementary perspectives, including entry-wise accuracy and structural preservation. The detailed formulations of these metrics are defined as follows.

\subsubsection{NMSE}
NMSE quantifies the relative prediction error by computing the ratio of the squared Frobenius norm of the estimation error to that of the ground truth, which is defined as
\vspace{-0.2cm}
\begin{align}
    {\rm NMSE} = \frac{\left\| \widehat{\Pm}^{\Delta}_{\Rcal} - \Pm^{\Delta}_{\Rcal} \right\|_{F}^2}{\left\| \Pm^{\Delta}_{\Rcal} \right\|_{F}^2},
\end{align}
where $\widehat{\Pm}^{\Delta}_{\Rcal}$ and $\Pm^{\Delta}_{\Rcal}$ denote the predicted and ground truth 3D RMs at resolution $\Delta$, respectively.

\subsubsection{RMSE}
RMSE measures the average magnitude of the prediction error by computing the square root of the mean squared difference between the predicted and ground truth values, which is defined as
\begin{align}
    {\rm RMSE} = \sqrt{\frac{1}{I_{\Delta}J_{\Delta}K_{\Delta}} \left\| \widehat{\Pm}^{\Delta}_{\Rcal} - \Pm^{\Delta}_{\Rcal} \right\|_F^2}.
\end{align}

\subsubsection{SSIM}
SSIM evaluates the perceptual similarity between the predicted and ground truth 3D RMs by jointly considering luminance, contrast, and structural information~\cite{SSIM}. In this paper, the SSIM is computed directly in the 3D domain using local statistics within cubic neighborhoods. For each grid index $\mathbf{p} = (i,j,k)$ in the 3D RM, $\forall i \in [I_{\Delta}], j \in [J_{\Delta}], k \in [K_{\Delta}]$, the local SSIM is defined as
\begin{align}
    \text{SSIM}(\mathbf{p}) = \frac{(2\mu_{\hat{x}}(\mathbf{p}) \mu_{x}(\mathbf{p}) + C_1)(2\sigma_{\hat{x}x}(\mathbf{p}) + C_2)}{(\mu_{\hat{x}}^2(\mathbf{p}) + \mu_{x}^2(\mathbf{p}) + C_1)(\sigma_{\hat{x}}^2(\mathbf{p}) + \sigma_{x}^2(\mathbf{p}) + C_2)},
\end{align}
where $\mu_{\hat{x}}(\mathbf{p})$ and $\mu_{x}(\mathbf{p})$ denote the local means of the predicted 3D RM $\widehat{\Pm}^{\Delta}_{\mathcal{R}}$ and the ground truth 3D RM $\Pm^{\Delta}_{\mathcal{R}}$, respectively. Similarly, $\sigma_{\hat{x}}^2(\mathbf{p})$ and $\sigma_{x}^2(\mathbf{p})$ are the local variances, and $\sigma_{\hat{x}x}(\mathbf{p})$ is the local covariance. The stabilization constants are $C_1 = (K_1 L)^2$ and $C_2 = (K_2 L)^2$, where $L$ is the dynamic range of the data and $K_1 = 0.01$, $K_2 = 0.03$ are default parameters. The final SSIM metric is obtained by averaging the local SSIM values over all valid grid locations:
\begin{align}
    \text{SSIM} = \frac{1}{|\Omega|} \sum_{\mathbf{p} \in \Omega} \text{SSIM}(\mathbf{p}),
\end{align}
where $\Omega$ denotes the set of interior grids excluding boundary regions affected by the filtering window.

\subsubsection{PSNR}
PSNR quantifies the reconstruction quality by measuring the ratio between the maximum possible signal power and the power of the reconstruction error in decibels, which is defined as
\begin{align}
    {\rm PSNR} = 10 \log_{10} \left( \frac{I_{\Delta}J_{\Delta}K_{\Delta}P_{\max}^2}{ \left\| \widehat{\Pm}^{\Delta}_{\Rcal} - \Pm^{\Delta}_{\Rcal} \right\|_F^2} \right),
\end{align}
where $P_{\max}$ denotes the maximum value of the ground truth 3D RM $\Pm^{\Delta}_{\Rcal}$.

\vspace{-0.45cm}
\subsection{Baseline Methods}
To comprehensively evaluate the effectiveness of each stage in the proposed framework, we consider the following baseline methods for comparison.
\begin{itemize}
    \item \textbf{RadioUNet3D-SR}: This baseline replaces the LR-Net in Stage~1 with RadioUNet3D, a 3D extension of the original RadioUNet~\cite{RadioUNet}, while retaining the SR-Net in Stage~2 for super-resolution. This comparison isolates the contribution of the proposed LR-Net architecture.
    
    \item \textbf{RadioUNet3D-Trilinear}: This baseline also employs RadioUNet3D in Stage~1 but substitutes the learned SR-Net in Stage~2 with conventional trilinear interpolation for upscaling. This comparison evaluates the benefit of learning-based super-resolution over classical interpolation when using RadioUNet3D.
    
    \item \textbf{LRNet-Trilinear}: This baseline utilizes the proposed LR-Net in Stage~1 to predict the low-resolution 3D RM but replaces the SR-Net in Stage~2 with trilinear interpolation. This comparison assesses the advantage of the learned super-resolution network over traditional interpolation within our framework.
\end{itemize}
\vspace{-0.2cm}
\begin{remark}
A natural baseline for evaluating the proposed two-stage architecture is a single-stage network that directly predicts the high-resolution 3D RM from the building layout and transmitter location, trained on a dataset where all $N$ environments have the corresponding high-resolution 3D RMs. However, our preliminary experiments reveal that this approach is computationally prohibitive and suffers from poor convergence with unsatisfactory performance. We attribute this to the excessively large output space of the high-resolution 3D RM, which impedes stable learning under our hardware constraints. Consequently, we exclude this baseline from our formal comparison. Nevertheless, these limitations further highlight the novelty of the proposed two-stage framework: by decomposing the problem into a coarse prediction stage followed by a learned super-resolution stage, our approach not only alleviates the computational burden but also achieves superior reconstruction quality.
\end{remark}

\vspace{-0.2cm}
\subsection{Quantitative Comparison}
\begin{figure*}[!t]
\centerline{\includegraphics[width=0.8\textwidth]{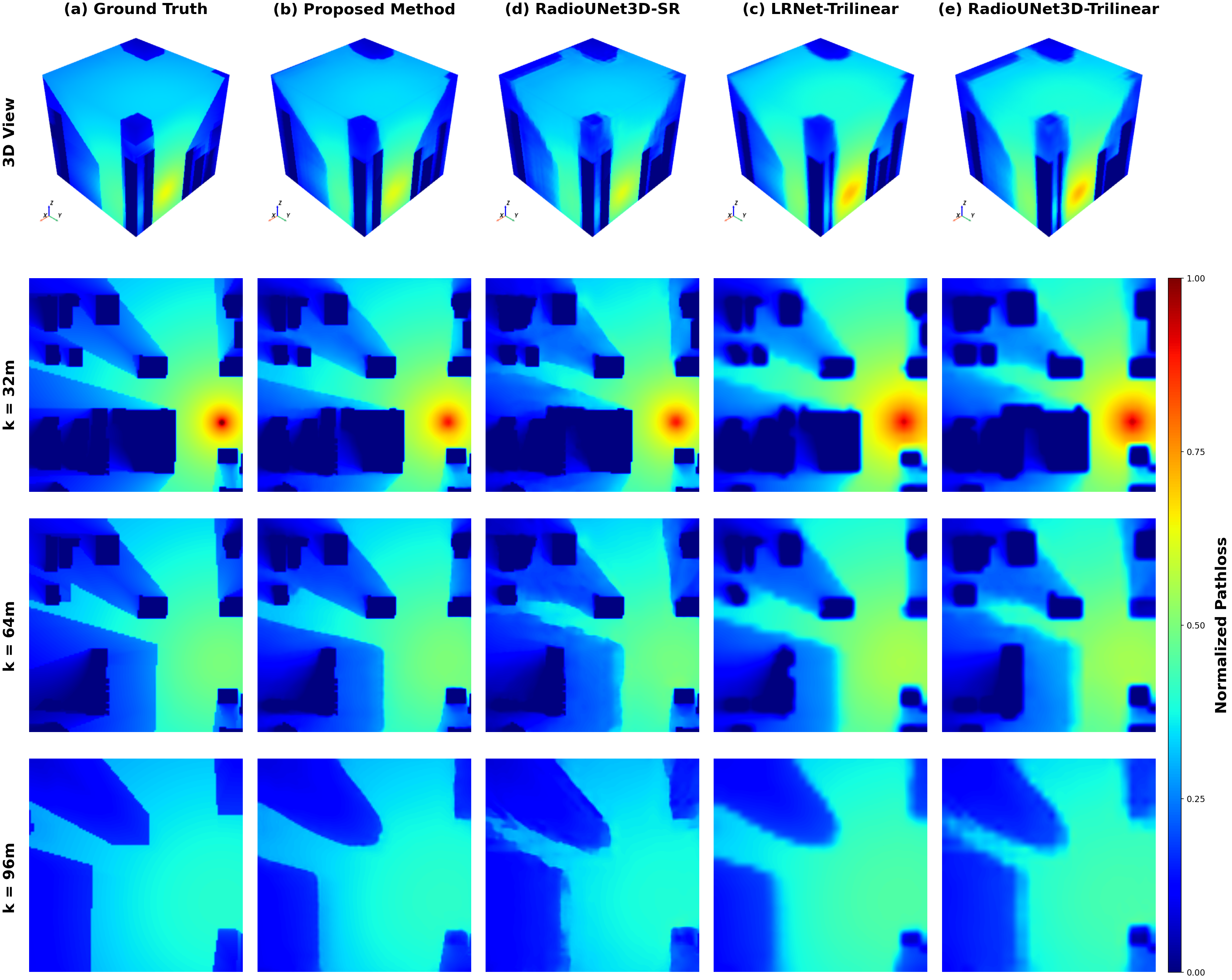}}
\vspace{-0.3cm}
\caption{Visual comparison of high-resolution 3D RMs: (a) Ground Truth, (b) Proposed Method, (c) RadioUNet3D-SR, (d) LRNet-Trilinear, and (e) RadioUNet3D-Trilinear. The top row shows the 3D volumetric view, while the bottom three rows display horizontal slices at altitudes $k = 32$, $64$, and $96$~m, respectively.}
\label{Fig: HR-RM Visualization}
\vspace{-0.5cm}
\end{figure*}

Table~\ref{Tab: Numerical Metrics} presents a quantitative comparison of the proposed method against the baseline methods across all four metrics. Several key observations can be drawn from this table. First, the proposed method achieves the best performance among all the approaches. Second, the comparison between the proposed method and RadioUNet3D-SR reveals that the proposed LR-Net in Stage~1 also contributes to performance improvement. Third, comparing RadioUNet3D-SR with RadioUNet3D-Trilinear, and the proposed method with LRNet-Trilinear, we observe that the learned SR-Net in Stage~2 substantially outperforms conventional trilinear interpolation.  These results collectively demonstrate each stage of the proposed framework plays an important role, and their integration enables accurate high-resolution 3D RM construction even with limited high-resolution training samples.
\begin{table}[t]
    \centering
    \caption{Performance Comparison in Terms of Numerical Metrics}
    \vspace{-0.2cm}
    \label{Tab: Numerical Metrics}
    \renewcommand{\arraystretch}{1.2}
    \begin{tabular}{l|c|c|c|c}
        \hline\hline
        Method & NMSE $\downarrow$ & RMSE $\downarrow$ & SSIM $\uparrow$ & PSNR $\uparrow$ \\
        \hline\hline
        RadioUNet3D-Trilinear & 0.0415 & 0.0537 & 0.8435 & 25.52 \\
        RadioUNet3D-SR & 0.0175 & 0.0298 & 0.9004 & 31.33 \\
        LRNet-Trilinear & 0.0420 & 0.0537 & 0.8642 & 25.51 \\
        \textbf{Proposed Method} & \textbf{0.0131} & \textbf{0.0260} & \textbf{0.9230} & \textbf{32.31} \\
        \hline\hline
    \end{tabular}
    \vspace{-0.4cm}
\end{table}

\vspace{-0.4cm}
\subsection{Visualization}
To provide intuitive insights into the reconstruction quality, we present visual comparisons of the high-resolution 3D RMs generated by different methods. Fig.~\ref{Fig: HR-RM Visualization} illustrates the 3D volumetric view of the 3D RMs along with 2D horizontal slices at three representative altitudes, i.e., $k = 32$, $64$, and $96$~m. As shown in Fig.~\ref{Fig: HR-RM Visualization}, the RM predicted by the proposed method closely resembles the ground truth, preserving both the overall signal distribution and fine-grained propagation characteristics. In particular, the proposed method accurately delineates the boundaries between line-of-sight and non-line-of-sight regions, which is essential for accurate coverage prediction in urban environments. In contrast, the baseline methods, especially those relying on trilinear interpolation, tend to produce overly smooth results that blur these critical boundaries.

\vspace{-0.4cm}
\subsection{Complexity Analysis} \label{Sec: Complexity Analysis}
To evaluate the deployment feasibility of the proposed method, we conduct a complexity analysis against the baseline approaches. Specifically, we measure the number of floating-point operations (FLOPs), inference time, and GPU memory consumption required to generate a single high-resolution 3D RM. In this experiment, we introduce an additional baseline, termed \emph{Single-Stage}, which adopts a similar architecture to the LR-Net in Stage~1 but directly maps the 3D EM $\Bm^{\Delta}$ and transmitter location map $\Qm^{\Delta}$ to the high-resolution 3D RM $\Pm_{\Rcal}^{\Delta}$ without the two-stage decomposition. As summarized in Table~\ref{Tab: Complexity}, the Single-Stage baseline incurs the highest computational cost across all three metrics, which consumes approximately four times more resources compared to the proposed method. This substantial resource demand stems from directly processing the high-dimensional output space without exploiting the hierarchical structure inherent in the hybrid dataset. Furthermore, by comparing RadioUNet3D-SR with RadioUNet3D-Trilinear, and the proposed method with LRNet-Trilinear, we observe that the computational bottleneck of DF-3DRME lies in the super-resolution operation in Stage~2. Nevertheless, with the significant performance gains demonstrated in Table~\ref{Tab: Numerical Metrics} and Fig.~\ref{Fig: HR-RM Visualization}, the additional computational overhead is well justified. Therefore, we believe the proposed method achieves a favorable trade-off between computational efficiency and reconstruction accuracy.
\begin{table}[t]
    \centering
    \caption{Complexity Comparison of Different Methods for Generating A Single High-Resolution 3D RM}
    \vspace{-0.2cm}
    \label{Tab: Complexity}
    \renewcommand{\arraystretch}{1.2}
    \begin{tabular}{l|c|c|c}
        \hline\hline
        \multirow{2}{*}{Method} & \multirow{2}{*}{FLOPs (G)} & Inference & GPU \\
         & & Time (ms) & Memory (MB) \\
        \hline\hline
        Single-Stage & 2820.29 & 454.33 & 6425.20 \\
        RadioUNet3D-Trilinear & 12.85 & 1.67 & 1480.22 \\
        RadioUNet3D-SR & 705.14 & 150.11 & 1557.25 \\
        LRNet-Trilinear & 10.49 & 6.23 & 71.46 \\
        \textbf{Proposed Method} & \textbf{702.78} & \textbf{156.94} & \textbf{1470.05} \\
        \hline\hline
    \end{tabular}
    \vspace{-0.4cm}
\end{table}

\vspace{-0.4cm}
\subsection{Impact of High-Resolution Data Availability} \label{Sec: HR Data Availability}
\begin{figure}[!t]
\centerline{\includegraphics[width=0.45\textwidth]{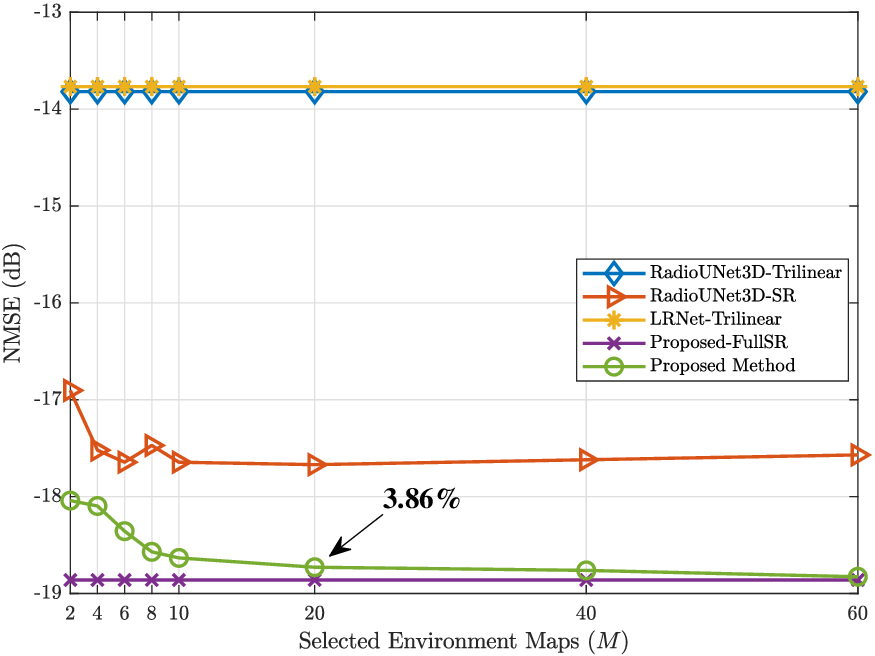}}
\vspace{-0.3cm}
\caption{NMSE v.s. the number of high-resolution training environments $M$.}
\label{Fig: NMSE vs M}
\vspace{-0.5cm}
\end{figure}

This subsection investigates how the number of high-resolution training samples affects the reconstruction performance of the proposed DF-3DRME. Recall that our hybrid dataset comprises $N = 517$ urban environments, each associated with a corresponding low-resolution 3D RM. Among these, only $M$ environments are paired with high-resolution 3D RM labels. To examine the reconstruction performance as a function of $M$, we vary $M \in \{2, 4, 6, 8, 10, 20, 40, 60\}$. In this experiment, we introduce an additional baseline, denoted as \emph{Proposed-FullSR}, which employs the same two-stage architecture as the proposed method but assumes that all $N = 517$ environments have the corresponding high-resolution RM labels during training. This baseline serves as an upper bound that represents the best achievable performance when sufficient high-resolution labels are available. Fig.~\ref{Fig: NMSE vs M} presents the NMSE performance in dB scale as a function of $M$. First, it is observed that both super-resolution methods significantly outperform their trilinear interpolation counterparts across all values of $M$. Second, the proposed method demonstrates remarkable sample efficiency. When $M = 20$, constituting only $3.86\%$ of all environments, the proposed method closely approaches the upper bound established by Proposed-FullSR. This result provides strong empirical validation that the proposed DF-3DRME exhibits data-friendly nature.

\vspace{-0.4cm}
\subsection{Impact of Low-Resolution Grid Size}
\begin{figure}[!t]
\centerline{\includegraphics[width=0.45\textwidth]{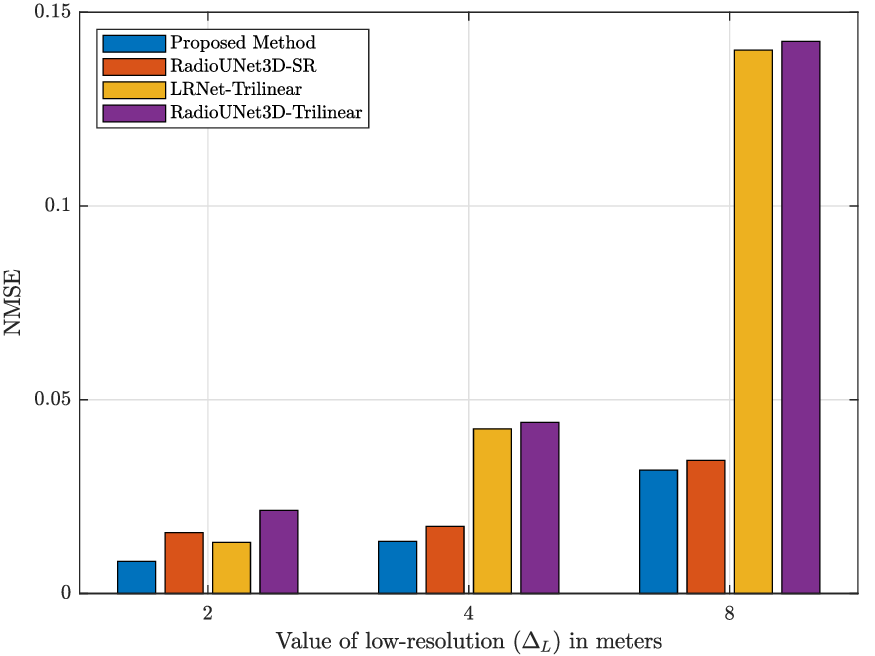}}
\vspace{-0.3cm}
\caption{NMSE v.s. the value of low-resolution grid size $\Delta_L$.}
\label{Fig: NMSE vs Delta_L}
\vspace{-0.3cm}
\end{figure}

This subsection investigates how the grid size of the low-resolution 3D RM, i.e., $\Delta_L$, affects the final reconstruction quality of the high-resolution 3D RM. We fix the target high-resolution grid size as $\Delta = 1$ and vary the low-resolution grid size as $\Delta_L \in \{2, 4, 8\}$. To isolate the impact of resolution coarseness on reconstruction performance, we assume that all $N = 517$ environments in the dataset have corresponding low-resolution 3D RMs available for each value of $\Delta_L$. Fig.~\ref{Fig: NMSE vs Delta_L} illustrates the NMSE performance in linear scale as a function of $\Delta_L$ for the proposed method and the baselines. It is observed that as $\Delta_L$ increases, the NMSE monotonically increases for all methods. This behavior is expected, because larger grid sizes inherently discard finer spatial details that cannot be fully recovered even with learned super-resolution. This finding also reveals a fundamental trade-off in practical deployment. On one hand, adopting a finer grid (smaller $\Delta_L$) preserves more spatial information and yields superior reconstruction accuracy, but at the expense of increased data collection efforts. On the other hand, employing a coarser grid (larger $\Delta_L$) reduces the cost of dataset construction but sacrifices reconstruction fidelity. Based on these results, $\Delta_L = 4$ is shown to be a favorable compromise, achieving a balanced trade-off between data acquisition overhead and reconstruction performance.

\vspace{-0.2cm}
\section{Conclusion} \label{Sec: Conclusion}
In this paper, we addressed the challenge of constructing high-resolution 3D RMs in practical scenarios where obtaining sufficient high-resolution training data is prohibitively expensive. We proposed a data-friendly DL framework termed DF-3DRME that strategically leverages a hybrid dataset structure, in which the majority of 3D EMs contain only low-resolution 3D RMs, while a small fraction of 3D EMs include both low-resolution and high-resolution 3D RMs. By decoupling the learning process into low-resolution 3D RM construction followed by super-resolution enhancement, our framework fully exploited the heterogeneous data at different resolutions. Extensive experimental results demonstrated that the proposed DF-3DRME significantly outperformed the baseline approaches in both numerical metrics and visual quality. More importantly, our framework achieved compelling reconstruction performance even with very limited high-resolution 3D RM labels. Future work will explore investigating different backbones for the implementation of Stage 1 and Stage 2 and extending this framework to dynamic environments.

\ifCLASSOPTIONcaptionsoff
  \newpage
\fi

%\begin{thebibliography}{1}
\vspace{-0.2cm}
\bibliographystyle{IEEEtran}
\bibliography{IEEEabrv.bib, reference.bib}

%\end{thebibliography}

\end{document}